\documentclass[conference]{IEEEtran}
\ifCLASSINFOpdf
   \usepackage[pdftex]{graphicx}
   \DeclareGraphicsExtensions{.pdf,.jpeg,.png}
\else
\fi

\usepackage{amsmath}
\usepackage{theorem}
\usepackage{fixmath}
\begin{document}
%

\title{Analytical Model of Proportional Fair Scheduling in Interference-limited OFDMA/LTE Networks}

\author{\IEEEauthorblockN{Donald Parruca, Marius Grysla, Simon G\"ortzen and James Gross}
\IEEEauthorblockA{UMIC Research Center\\RWTH Aachen University, Germany\\
Email: \{parruca, goertzen, gross\}@umic.rwth-aachen.de \\ marius.grysla@rwth-aachen.de}
}


%


\maketitle

\begin{abstract}
Various system tasks like interference coordination, handover decisions, admission control etc. in upcoming cellular networks require precise mid-term (spanning over a few seconds) performance models. Due to channel-dependent scheduling at the base station, these performance models are not simple to obtain. Furthermore, upcoming cellular systems will be interference-limited, hence, the way interference is modeled is crucial for the accuracy. In this paper we present an analytical model for the SINR distribution of the \textit{scheduled} subcarriers of an OFDMA system with proportional fair scheduling. The model takes the precise SINR distribution into account. We furthermore refine our model with respect to uniform modulation and coding, as applied in LTE networks. The derived models are validated by means of simulations. In additon, we show that our models are approximate estimators for the performance of rate-based proportional fair scheduling, while they outperform some simpler prediction models from related work significantly.\footnote{This work has been submitted to the IEEE for possible publication. Copyright may be transferred without notice, after which this version may no longer be accessible}

\end{abstract}

\IEEEpeerreviewmaketitle



\section{Introduction}
\label{sec:intro}
Cellular networks are facing a considerable increase in mobile data traffic over the next ten years \cite{trafficForecast2010}. The current traffic impact generated from smart phones is to some extent evidence for the development still ahead. To satisfy these demands, ambitious requirements are set for the upcoming 4-th generation of cellular networks like IEEE 802.16 \cite{WimaxStandard} and 3GPPs LTE (long term evolution) \cite{3GPP_TS36_321}.  The demand of higher capacity on the one hand and the presence of scarce frequency resources on the other, poses challenging aspects on their system design. Frequency reuse of one is applied for a flexible usage of the resource allocation especially for asymmetrical load distributions. Additionally, feedback based resource allocation schemes (scheduling) are introduced for an efficient exploitation of the available resources. 

The scheduling algorithm running in the base station for the down-link can quickly adapt to fading channel conditions thanks to the feedback of the channel states sent from the mobile stations. Opportunistic schedulers~\cite{Jalali2000} assign the resources to the mobile terminals having the best channel conditions by taking advantage of multi-user diversity. Although it delivers the best overall system throughput, terminals with bad channel conditions over an extended period of time experience significant QoS degradation as they are less likely to be scheduled at all. In order to resolve this problem, proportional fair scheduling (PFS) has been proposed. For this the instantaneous channel states (either characterized by the rate or by the signal-to-interference-and-noise ratio) are scaled by dividing with the average channel state. Based on this modified channel characterization, the scheduler opportunistically assigns the resources. Therefore, the resources end up with the terminals that have the relatively best channel states in comparison to their average state. Due to the characteristic of PFS  to support terminals more evenly, it is widely accepted and used as reference scheduler in many other investigations~\cite{Zhenning2011,Qian2012,Huang2010}.

Despite the beneficial characteristics of PFS, it obviously makes system analysis tougher. This is due to the fact that the PFS algorithm 'transforms' the stochastic properties of the resources assigned~\cite{gross09}. For instance, assuming the channel states to be exponentially distributed (which is a standard model for the signal-to-noise ratio), the resulting distribution of the channel states of the \textit{assigned} resources are clearly modified in a beneficial way. However, the exact impact is hard to derive while for system tasks like admission control, handover, interference coordination and load balancing a precise model is clearly of interest. This has lead to multiple attempts to describe the impact of PFS analytically. For instance, in~\cite{Caire2007} the authors derive upper and lower bounds for the spectral efficiency of PFS in noise-limited cases. The work was extended in \cite{Park2008} to interference-limited systems. Nevertheless, these works were not able to provide an exact analytical expression. Further attempts have been made in~\cite{Choi2007,liu2008,liu2010,Almatarneh2010PFS,ahmed2012analytical}. However, all these works relied on the assumption that the instantaneous scheduled rates are Gaussian-distributed under PFS. While this assumption might hold for noise-limited systems, it clearly does not for interference-limited ones. Hence, these analytical models can not be applied if dealing with interference-limited scenarios. Nevertheless, these scenarios are the most interesting in upcoming fourth generation systems, as due to the frequency reuse of one, the systems are mostly interference-limited. 


In this work, we concentrate on an exact analytical model of PFS specifically considering interference-limited OFDMA/LTE systems. We first derive the distribution of the transformed SINR channel states when PFS resource allocation is performed. This basic expression takes perfect channel state information at the base station into account and considers also a flexible allocation of modulation and coding schemes to resources. After validating our analysis, we then turn to additionally modeling impairments that arise in LTE systems. Namely, the channel state feedback is delayed and quantized while modulation and coding is applied uniformly over all resources assigned to a terminal during one slot. We show that our novel model is quite exact even when limited channel-feedback and fixed modulation/coding schemes are considered. Moreover, we show that all other presented analytical models lead for the interference-limited case to much bigger errors in performance prediction. This is important as any control algorithm (handover, load balancing, interference coordination) implemented on top will lead to poor performance due to the estimation error in the PFS performance model. To the best of our knowledge, our model is novel with respect to related work.

Our remaining work is structured in the following way. In Section~\ref{sec:model} we introduce the system model and define the problem that we are interested in. Section~\ref{sec:basic_analysis} presents the major analysis of our work while some refinements for LTE are presented in Section~\ref{sec:practical_aspects}. We finally compare our prediction model with some other ones from related work in Section~\ref{sec:model_comp} before we conclude the paper in Section~\ref{sec:conclusions}.

\section{Basic System Model}
\label{sec:model}
Initially, we consider the down-link of a cellular (LTE-like) OFDMA system. Time is split up into transmission slots with index $t$. The slots are also referred to as Transmission Time Intervals (TTI) in the LTE context. They have a duration of $T_{\mathrm{TTI}}$ seconds. Orthogonal frequency division multiplexing (OFDM) is the basic transmission scheme. Hence, the system bandwidth $B$ is split into many subcarriers out of which $R$ adjacent ones are always bundled together for resource allocation. We refer to these bundles as resource blocks and assume that there are $N$ such resource blocks in the system. A resource block is the smallest frequency unit for resource allocation at the base station. Furthermore, per slot (i.e. per TTI) each subcarrier can convey $S$ symbols to some receiver. Therefore, a resource block carries $R \cdot S$ resource elements in total. In the following we focus on a particular cell of a cellular deployment. There are $J$ terminals (also referred to as user equipment - UE - in LTE) currently in this cell and we denote the set of these terminals by $\mathcal{J}$. Each of these terminals is also subject to interfering down-link signals from the neighboring cells. 

The communication in each cell is centrally controlled by the base station (also referred to as eNB). Per TTI the base station assigns different resource blocks to different terminals depending on the channel states of the terminals as well as the currently backlogged data. The down-link channel state for terminal $j$ on resource block $n$ at slot $t$ is given by the signal-to-interference-and-noise ratio (SINR) $\gamma_{j,n}\left(t\right)$ defined as:
\begin{equation}
  \label{eq:sinr_definition}
  \gamma_{j,n}\left(t\right) = \frac{p_n^s g_{j,n}^s(t)}{p_n^i g_{j,n}^i(t) + \eta} \; \mathrm{.}
\end{equation}
Here, $p_n^s$ and $p_n^i$ are the transmission powers for the signal of interest and of the interfering signal. Note that these can take arbitrary values. We consider in the following one dominant interfering signal for each terminal. Furthermore, $g_{j,n}^s(t)$ and $g_{j,n}^i(t)$ denote the instantaneous channel gains for the signal of interest and for the interfering signal. Finally, $\eta$ denotes the noise power.

Given the SINR on resource block $n$, we denote by $C\left(\gamma_{j,n}\right)$ the spectral efficiency of the modulation and coding scheme used to transmit resource block $n$ regrading terminal $j$. Example functions can be found in~\cite{3gpp_ts36.216,Ikuno2010}, alternatively the Shannon formula $\log_2\left(1+\gamma\right)$ is often used for this as theoretical upper bound. Note that we initially assume $C$ to be independent from any other resource block, i.e. the modulation and coding type for $n$ can be chosen independent of the other resource blocks. We will refine this assumption later in Section~\ref{sec:practical_aspects}.

The considered resource allocation scheme is the proportional fair scheduling (PFS) algorithm. Given the SINR values $\gamma_{j,n}$ the scheduler operates as follows. Let: 
\begin{equation}
\bar{\gamma}_{j,n}\left(t\right)=\frac{1}{W}\sum_{i=t-W}^t \gamma_{j,n}\left(i\right)
\end{equation}
express the average SINR during the last $W$ time slots on resource block $n$ for UE $j$. $W$ is also known as the time window of proportional fair scheduling. Then $\widehat{\gamma}_{j,n}(t)$ denotes the scaled, instantaneous SINR defined as:
\begin{equation}
\widehat{\gamma}_{j,n}(t)=\frac{\gamma_{j,n}(t)}{\bar{\gamma}_{j,n}(t)} \; \mathrm{.}
\label{eqn:PFS_operation}
\end{equation}
The PFS algorithm allocates resource block $n$ to terminal $j^{*}$ in case it has the best scaled SINR among the other users. In other words, every TTI the following allocations are done:
\begin{equation}
\forall n: \; j^{*}_n\left(t\right)=\arg \max_{j\in \mathcal{J}} \widehat{\gamma}_{j,n}(t) \; \mathrm{.}
\end{equation}
In order to perform this algorithm, the scheduler requires the channel states $\gamma_{j,n}$. We initially assume that this information is accurately provided by the system. Later on in Section~\ref{sec:practical_aspects}, we will also consider consequences from the implementation of the feedback channel in LTE systems. 

We will concentrate in this work on formulating a mathematical model which describes the impact of PFS on the distribution of the SINR of the \textit{scheduled} resource blocks for some terminal. From this we can easily derive for instance the average SINR or the average obtained rate per terminal per cell. Once obtaining these characterization for an ideal OFDMA system with perfect channel state information, we will also consider the impact due to uniform modulation and coding schemes as well as due to quantized feedback as they are implemented in LTE networks.

\section{Basic Analysis of PFS in Interference-Limited Scenarios}\label{sec:basic_analysis}

The temporary channel gains $g_{j,n}$ of the signal of interest as well as of the interfering signal are modeled as exponentially distributed random variables (RV) according to a Rayleigh fading model. We assume the average channel gains $\overline{g}_{j,n}$ to be stationary for some longer duration. Typically, this is the case for a few hundred milliseconds up to seconds. Then the SINR $\gamma_{j,n}(t)$ on resource block $n$ for terminal $j$ is a random variable denoted by $X_{j,n}$. The corresponding density function $f_{X_{j,n}}\left(x\right)$ of the SINR is well known to be~\cite{gross2010}:
\begin{align}
f_{X_{j,n}}\left(x\right)  &=&  \left[\frac{\eta}{P_{j,n}^i \cdot x_{j,n} + P_{j,n}^s} + \frac{P_{j,n}^s \cdot P_{j,n}^i}{(P_{j,n}^i \cdot x+ P_{j,n}^s)^2}\right] \nonumber \\ &&\cdot \exp \left( - \frac{\eta}{P_{j,n}^s} x\right) , \label{eqn:base_pdf}
\end{align}
with the corresponding cumulative distribution function:
\begin{align}
F_{X_{j,n}}\left(x\right)  = 1- \frac{P_{j,n}^s}{(P_{j,n}^i \cdot x+ P_{j,n}^s)}\cdot \exp \left( - \frac{\eta}{P_{j,n}^s} x\right)
\end{align} where $\mathnormal{P_{j,n}^s = p_n^s \overline{g}_{j,n}^s, P_{j,n}^i = p_n^i \overline{g}_{j,n}^i}$ are the average received powers of the signal of interest and of the interfering signal.

For an infinite window size $W$ we have:
\begin{eqnarray}
\mathnormal{\bar{\gamma}_{j,n}} & = & \lim_{W\rightarrow \infty}\left(\frac{1}{W}\sum_{i=t-W}^t \gamma_{j,n}(i) \right)\nonumber \\
&=& \text{E}[X_{j,n}] \nonumber \\
&=& \int_0^\infty x \cdot f_{X_{j,n}}\left(x\right) dx \nonumber \\
&=&\int_{0}^{+\infty} 
x \cdot \mathnormal{\left[\frac{\eta}{P_{j,n}^i \cdot x + P_{j,n}^s} +
\frac{P_{j,n}^s \cdot P_{j,n}^i}{\left(P_{j,n}^i \cdot x + P_{j,n}^s\right)^2}\right]} \ \ \ \ \ \ \nonumber  \\  &&
   \mathnormal{\cdot \exp \left(-\frac{\eta}{P_{j,n}^s} \cdot x\right)~dx} \nonumber \\ 
&=&\frac{P_{j,n}^s \cdot \textrm{Ei}\left( -\frac{\eta x}{P_{j,n}^s} - \frac{\eta}{P_{j,n}^i} \right)}{P_{j,n}^i} \cdot \exp\left(\frac{\eta}{P_{j,n}^i} \right) \nonumber \\ &&
+\left( \frac{{\left(P_{j,n}^s\right)}^2}{P_{j,n}^i(P_{j,n}^ix + P_{j,n}^s)} - \frac{P_{j,n}^s}{P_{j,n}^i} \right) \cdot 
  \exp \left(- \frac{\eta x}{P_{j,n}^s} \right)   
     \mathrm{.}
\label{static}
\end{eqnarray}
Therein Ei denotes the exponential integral. Equation~\eqref{static} is also referred to as static SINR estimation in case RB $n$ is persistently assigned to UE $j$ \cite{Parruca2013}. Then, following rules of random variable transformation \cite{papoulis:_Probability} the distribution function $F_{\widehat{X}_{j,n}}(x)$  and the PDF $f_{\widehat{X}_{j,n}}(x)$ of the random variable $\widehat{X}_{j,n} = \frac{X_{j,n}}{\text{E}[X_{j,n}] }$ representing the scaled SINR can be calculated:
\begin{equation}
F_{\widehat{X}_{j,n}}(x)=F_{X_{j,n}}(\text{E}[X_{j,n}] \cdot x)
\end{equation}
\begin{equation}
f_{\widehat{X}_{j,n}}(x)=\text{E}[X_{j,n}]\cdot f_{X_{j,n}}(\text{E}[X_{j,n}]\cdot x)
\end{equation}

Let us define the scheduling indicator $M_{j,n}$ as a new binary random variable which will help in analyzing the PF scheduler. It is one whenever terminal $j$ is scheduled the resource block $n$. This happens when it has the highest scaled SINR $\widehat{X}_{j,n}$ on RB $n$ among all other terminals. Likewise, the indicator is zero if terminal $j$ is not scheduled.
Given the scheduling indicator $M_{j,n}$, we can now denote the density function of the random SINR in case that this resource block is scheduled for terminal $j$ as $f_{X_{j,n}|M_{j,n} = 1}(x)$. Assuming we have an expression for this density, the expected rate $\mathcal{R}_{j,n}$ on resource block $n$ for terminal $j$ can be obtained as:
\begin{align}
  \label{eq:expected_rate_rb}
  \mathcal{R}_{j,n} &=& \frac{R \cdot S}{T_{\text{TTI}}}\int_0^{\infty} C(x) \cdot f_{X_{j,n}| M_{j,n} = 1}(x) ~ dx \nonumber \\ && \cdot \text{P}\left(M_{j,n} = 1\right) \; \mathrm{.} 
\end{align}
$\text{P}\left(M_{j,n} = 1\right)$ is the probability that terminal $j$ will be scheduled on RB $n$. Finally, the total rate per terminal is simply given by the sum of the expected rates per resource block: 
\begin{equation}
\mathcal{R}_{j}=\sum_{n=1}^{N} \mathcal{R}_{j,n}.
\label{eqn:totalRateIndepMCS}
\end{equation}

\subsection{Calculation of Scheduled SINR PDF}
To obtain the average rate, we need to derive the conditional density function $f_{X_{j,n} | M_{j,n} = 1}(x)$. From Bayes' theorem we have:
\begin{align}
  f_{X_{j,n}|M_{j,n} = 1}(x) 
  = \frac{f_{M_{j,n}=1|X_{j,n}}(x) \cdot f_{X_{j,n}}(x)}{\text{P}(M_{j,n}=1)} 
\end{align}
Then, assuming the scaled SINR random variables $\widehat{X}_{j,n}$ to be independent among the terminals $j$ and denoting the maximum order statistics among them by $\widehat{X}_{\forall i\neq j\in \mathcal{J}}^{\max}$ with distribution function $F_{\widehat{X}_{\forall i\neq j\in \mathcal{J}}^{\max}}(x)$, it follows:
\begin{align}	
\text{P}(M&_{j,n}=1) = P(\widehat{X}_{j,n} \geq \max_{\forall i\neq j\in \mathcal{J}}(\widehat{X}_{i,n})) \nonumber \\
   &= \int_0^{\infty} f_{\widehat{X}_{j,n}}(z)  \int_0^{z} f_{\widehat{X}_{\forall i\neq j\in \mathcal{J}}^{\max}}(y) ~ dy ~ dz \nonumber \\
  &= \int_0^{\infty} f_{\widehat{X}_{j,n}}(z) \cdot F_{\widehat{X}_{\forall i\neq j\in \mathcal{J}}^{\max}}(z) ~ dz \nonumber \\
  &= \int_0^{\infty} f_{\widehat{X}_{j,n}}(z) \cdot \prod_{\forall i\neq j\in \mathcal{J}} F_{\widehat{X}_{i,n}}(z) ~ dz  \nonumber \\
   &= \int_0^{\infty} f_{\widehat{X}_{j,n}}(z) \cdot \prod_{\forall i\neq j\in \mathcal{J}} F_{X_{i,n}}(\text{E}[X_{i,n}]\cdot z) ~ dz 
   \label{eq:scheduling_probability_last}
\end{align}
The only unknown term left for the transformed PDF calculation is the term $f_{M_{j,n} = 1 | X_{j,n}}(x)$. It can be simplified by rewriting it to a function containing the best-order statistics CDF:
\begin{align}
  f_{M_{j,n} = 1 | X_{j,n}}(x) &= \text{P}(M_{j,n} = 1 | X_{j,n} = x)  \nonumber \\ 
  &=\text{P}\left(\widehat{X}_{j,n} \geq \max_{\forall i\neq j\in \mathcal{J}}(\widehat{X}_{i,n}) | X_{j,n} = x\right) \nonumber \\
  &= \text{P}\left(\frac{X_{j,n}}{\text{E}[X_{j,n}]} \geq \max_{\forall i\neq j\in \mathcal{J}}(\widehat{X}_{i,n}) | X_{j,n} = x\right) \nonumber \\
  &= \text{P}\left(\frac{x}{\text{E}[X_{j,n}]} \geq \max_{\forall i\neq j\in \mathcal{J}}(\widehat{X}_{i,n})\right) \nonumber \\
  &= F_{\widehat{X}_{\forall i\neq j\in \mathcal{J},n}^{\max}}\left(\frac{x}{\text{E}[X_{j,n}]}\right) \nonumber \\
  &= \prod_{\forall i\neq j\in \mathcal{J}} F_{\widehat{X}_{i,n}}\left(\frac{x}{\text{E}[X_{j,n}]}\right) \nonumber \\
  &= \prod_{\forall i\neq j\in \mathcal{J}} F_{X_{i,n}}\left(\frac{\text{E}[X_{i,n}]}{\text{E}[X_{j,n}]}  \cdot x\right)
\end{align}
Summing up the above, the probability density function of the scheduled SINR can be expressed as:
\begin{multline}\label{eqn:transformed_SINR}
  f_{X_{j,n}| M_{j,n}=1}(x) =\\
  \frac{\prod_{\forall i\neq j\in \mathcal{J}} F_{X_{i,n}}\left(\frac{\text{E}[{X}_{i,n}]}{\text{E}[X_{j,n}]}  \cdot x\right) \cdot f_{X_{j,n}}(x)}
  {\int_0^{\infty} f_{\widehat{X}_{j,n}}(z) \cdot \prod_{\forall i\neq j\in \mathcal{J}} F_{X_{i,n}}(\text{E}[X_{i,n}]\cdot z) ~ dz }
\end{multline}

By replacing $\text{E}[X_{j,n}]=1$ in the last formula, the PDF transformation of opportunistic scheduling can be obtained as well. In opportunistic scheduling the history of the channel states (see Formula \ref{eqn:PFS_operation}) is not considered for the RB assignment decisions. However, opportunistic scheduling is out of scope in this work and we concentrate further on PFS. Unfortunately, it is difficult to obtain a closed form solution of the integral in  the denominator of $f_{X_{j,n}| M_{j,n}=1}(x)$. Nevertheless, numerical methods can be used and relatively precise results can be obtained. The calculation of the $\text{P}(M_{j,n}=1)$, being the denominator in the last formula is a novelty compared to related work in \cite{liu2008,liu2010}. It gives a clear statement on the assignment probability of a MS to a RB, unlike \cite{Ma2007PFS} assuming each user has the same channel access probability. This will be used again later on in the analysis of practical system aspects.

\subsection{Model Validation}
\label{section:model_evaluation_indepMCS}
\begin{figure}[ht]
\centering
\includegraphics[width=3.3in]{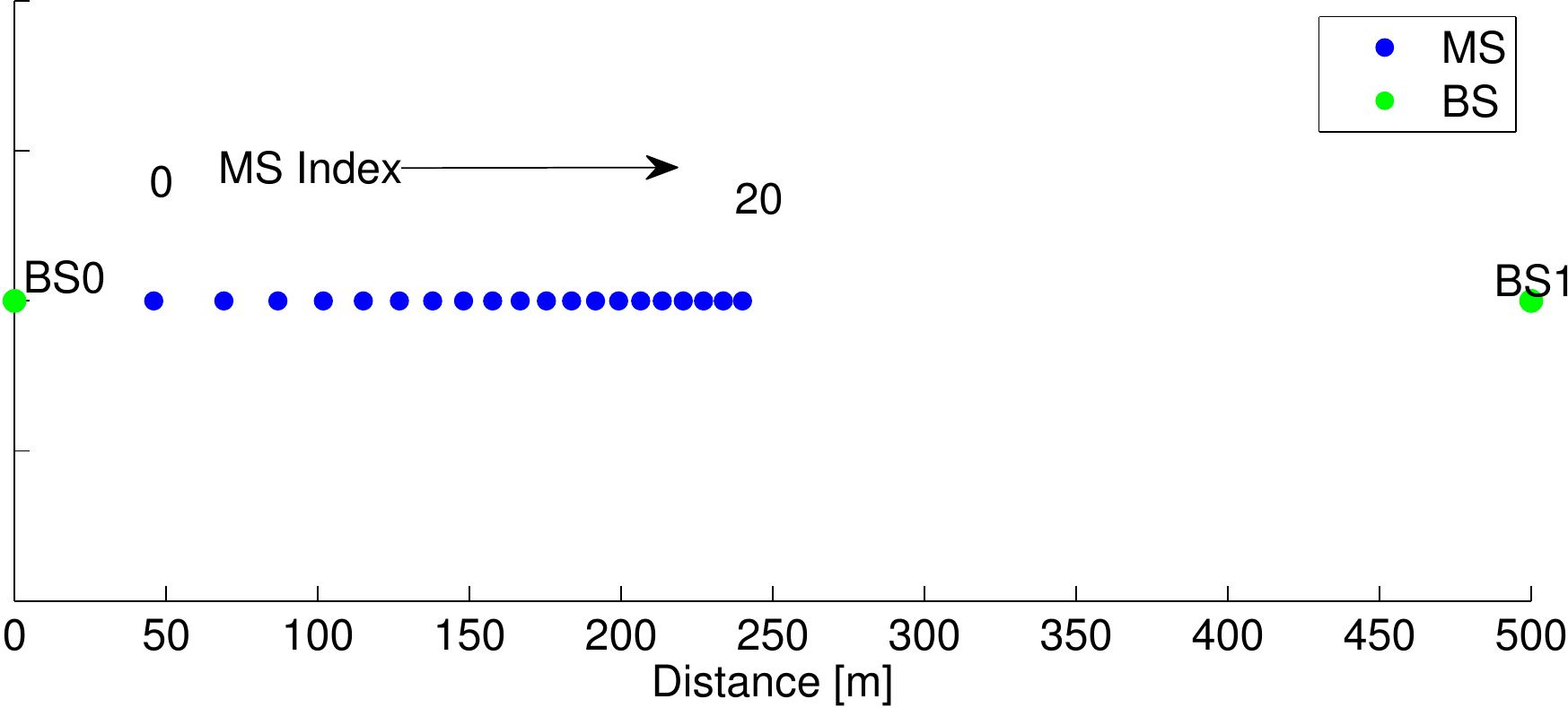}
\caption{MS location for model validation}
\label{fig:MS_distribution}
\end{figure}

In order to validate the theoretical model developed so far, we compare the derived rates with empirical results from system level simulations. We consider a simplified set up where 20 terminals are distributed along a line as shown in Figure~\ref{fig:MS_distribution}. The serving base station is to the left. The considered carrier frequency is set to $2$ GHz and the system has a bandwidth of $5$ MHz. Accordingly, we set the number of resource blocks to $N=25$ with $R=7$ symbols per resource block each composed of $S=12$ subcarriers. There is one interfering base station situated 500 m from the serving one (to the right in the figure). A standard path loss model from 3GPP is used (urban scenario) with $35.2+35\log_{10}(d)$ for the calculation of the average channel gains $\bar{g}_{j,n}^s$, $\bar{g}_{j,n}^i$ (where $d$ is the distance between terminals and base station in meters). Meanwhile, for the fast fading simulation the Jake's model is assumed. The total transmission power is $20$ W ($0.8$ W per RB), the noise power per resource block is set to $-112$ dBm. The simulated time equals $5$ seconds. The window size for the PFS is set to $100$. For the link to system mapping the SINR to spectral efficiency function $C(x)$ from \cite{Ikuno2010} was used and a full buffer traffic model was simulated.

\begin{figure}[ht]
\centering
\vspace{-5pt}
\includegraphics[width=3.3in]{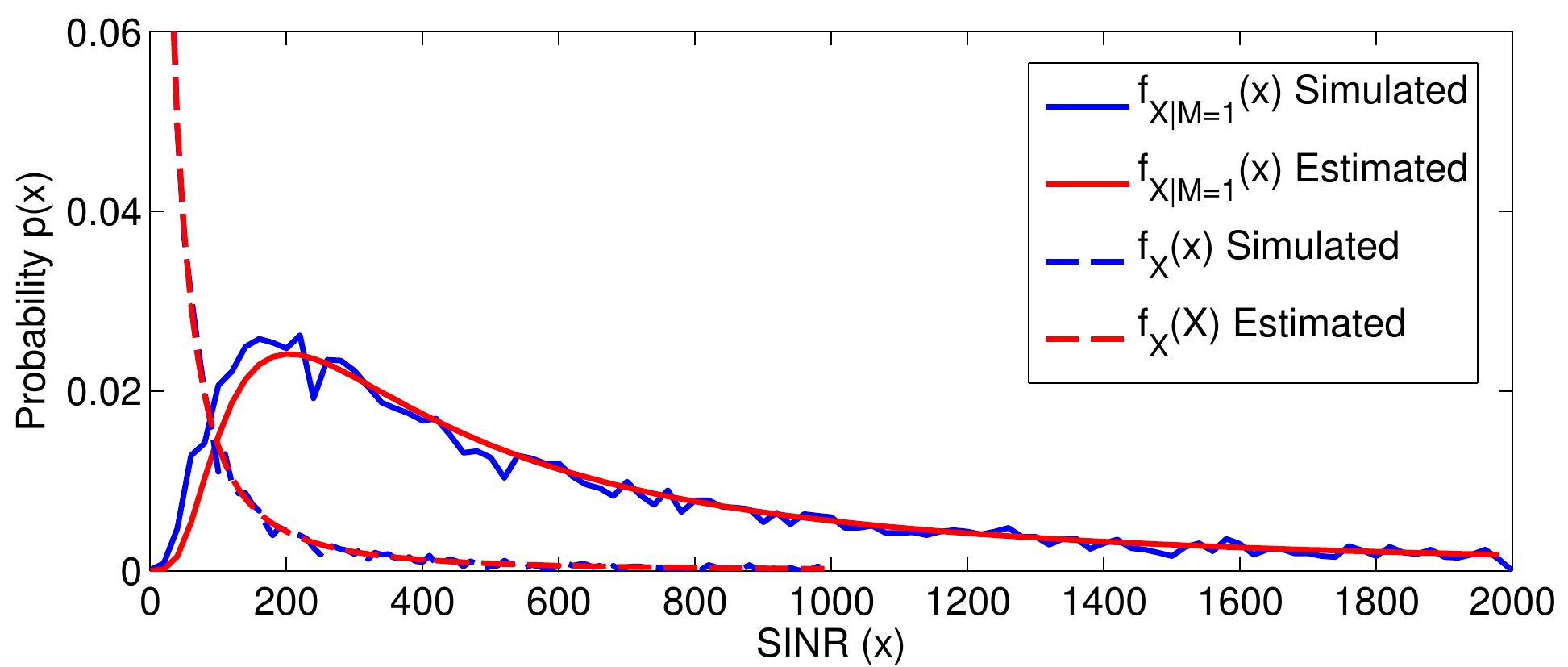}
\caption{Comparison of the estimated SINR PDF model vs. simulations}
\label{fig:sinr_transformation}
\end{figure}

\begin{figure}[ht!]
\centering
\vspace{-5pt}
\includegraphics[width=3.3in]{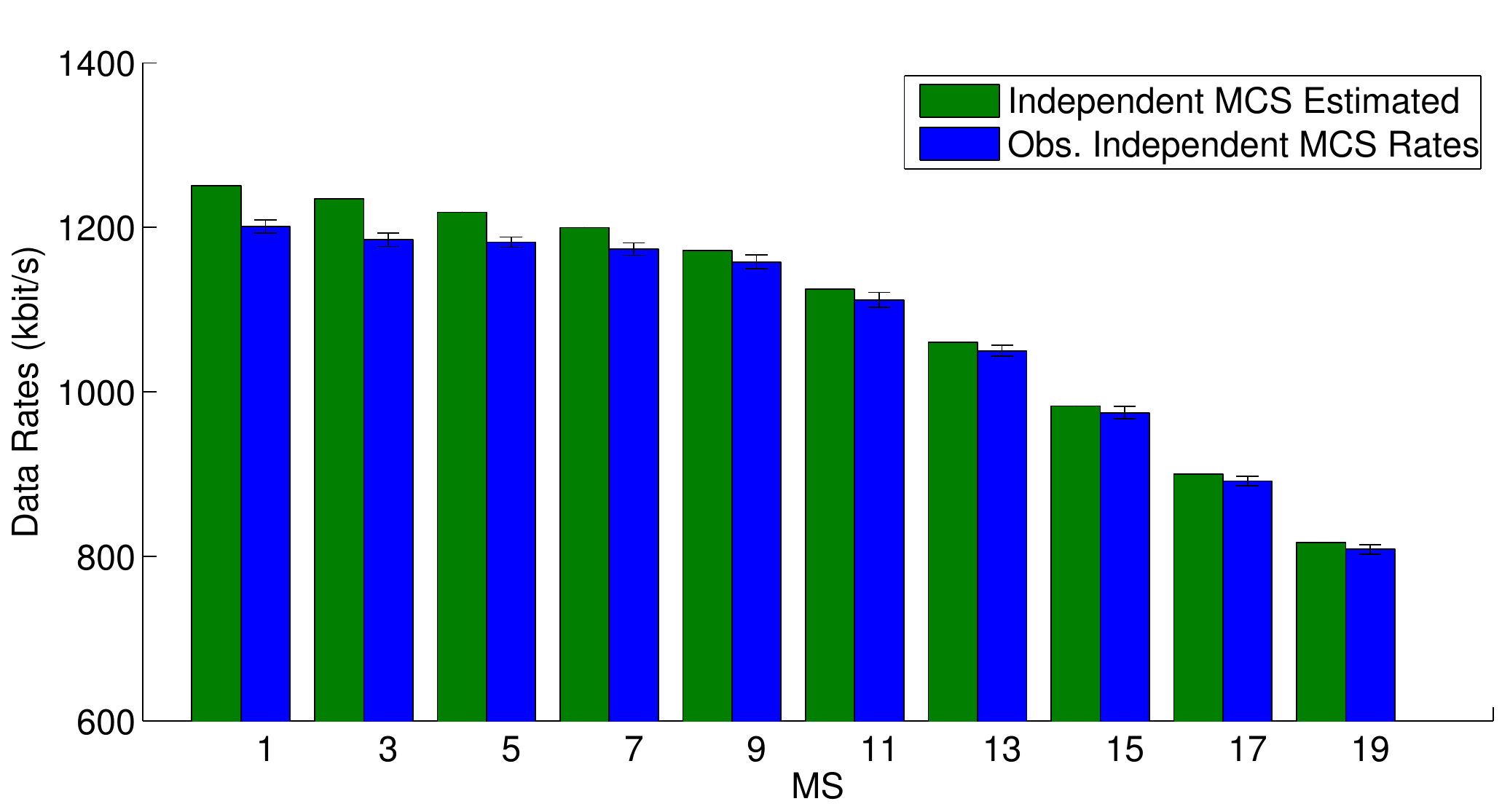}
\caption{Comparison of the estimated rates vs obtained rates from simulations}
\label{fig:indepMCS_vs_simulations}
\end{figure}

Figure~\ref{fig:sinr_transformation} shows the SINR PDF transformation of the basic SINR distribution $f_{X_{j,n}}(x)$ to the scheduled SINR $f_{X_{j,n}|M_{j,n}=1}(x)$ for one RB of one the MSs. As it can be noticed there is a very good match between the observed SINR distributions from simulations and the models presented in Formulas \ref{eqn:base_pdf} and \ref{eqn:transformed_SINR}. The simulations were repeated 30 times for different seeds of the fading process and the average rates per terminal compared with the model from Formula \ref{eqn:totalRateIndepMCS} are shown in Figure \ref{fig:indepMCS_vs_simulations}. Again, the comparison between simulations and the models presented in Formula \ref{eqn:totalRateIndepMCS} shows a good match. The small discrepancy noticed for the terminals 1-7 comes due to the finite size of the PFS window $W$, which in our theoretical model was assumed to be of infinite size.


\section{Practical Aspects of PF Scheduling}
\label{sec:practical_aspects}
Until now we have considered perfect channel state knowledge at the base station as well as modulation and coding schemes that can be set independently per resource block. These are rather optimistic assumptions as typical OFDMA realization are subject to some limitations. For instance, in LTE all resource blocks assigned to a certain terminal are utilized by the same modulation and coding type. Likewise, a quantized channel feedback is sent from the terminals to the base station. We discuss the impact of these two design decisions on our mathematical model in the following. 

\subsection{Common Modulation and Coding Scheme}
The common MCS restriction introduces additional refinements to the rate estimation model introduced so far. This is mainly due to the joint statistics of the currently assigned RBs. Now, the MCS selection will depend on the instantaneous channel realizations of the jointly assigned RBs. The way how the MCS will be selected is not standardized in LTE, nevertheless there are some common approaches used. In our case, we select a conservative MCS selection process based on the worst SINR realization of the assigned RBs. This requires us to make use of the minimum order statistics. For a nonempty subset $\mathcal{A} \subseteq \{1, \ldots, N \}$ of the RBs we define:
\begin{equation}
  M_{j,\mathcal{A}} = \left\{
    \begin{array}{ll}
      1, & \mathcal{A} = \{n\, |\, M_{j,n}=1 \},\\
      0, & \textrm{else.}
    \end{array}
  \right.
\end{equation}\label{def:M}
We model hence the transmission throughput to depend on the minimal SINR of the assigned RBs to $j$:
\begin{equation}
X_{j,\mathcal{A}}^{\min}=\min_{n \in \mathcal{A}} X_{j,n}
\end{equation}
Using this notation, we can express the rate for MS $j$ with a scheduled set of RBs $\mathcal{A}$ as:
\begin{equation}
   \label{eq:independent_mcs_rate}
   \mathcal{R}_{j,\mathcal{A}} = \text{P}_{j,\mathcal{A}} |\mathcal{A}|  \frac{R \cdot S}{T_{\text{TTI}}} \int_0^{\infty} f_{X_{j,\mathcal{A}}^{\min}| M_{j,\mathcal{A}}=1}(x) \cdot C(x) ~ dx
 \end{equation}
in which $|\mathcal{A}|$ is the size of the set $\mathcal{A}$ and $P_{j,\mathcal{A}} = \text{P}\left(M_{j,\mathcal{A}} = 1\right)$ is the probability that the set $\mathcal{A}$ will be assigned to MS $j$. The computation of this probability is involved as the resource blocks in $\mathcal{A}$ can be subject to correlation (due to the delay spread of the propagation environment). We continue our analysis under the assumption that the resource blocks of $\mathcal{A}$ are statistically independent. This introduces an error that we quantize afterwards in the validation part (finding that it is rather small). Therefore, we obtain for the probability:
\begin{equation}
\text{P}_{j,\mathcal{A}}= \prod_{n \in \mathcal{A}}\text{P}(M_{j,n}=1) \prod_{n \notin \mathcal{A}}[1-\text{P}(M_{j,n}=1)]
\end{equation}
The CDF of minimum order statistic~$X_{j,\mathcal{A}}^{\min}$ conditioned under~$M_{j,\mathcal{A}}=1$ computes to
\begin{IEEEeqnarray}{RL}
  F_{X_{j,\mathcal{A}}^{\min}|M_{j,\mathcal{A}}=1}(x) &= \text{P}\left( \min_{n\in\mathcal{A}} X_{j,n}\leq x \,|\, M_{j,\mathcal{A}}=1 \right)\nonumber \\[2pt]
  & = 1 - \text{P}\left( X_{j,n}>x \,\forall\,n\in\mathcal{A}\,|\, M_{j,\mathcal{A}}=1 \right)\nonumber \\[2pt]
  & = 1 - \prod_{n\in\mathcal{A}} \bar{F}_{X_{j,n}|M_{j,n}=1}(x),
\end{IEEEeqnarray}
in which~$\bar{F}_{X_{j,n}|M_{j,n}=1}(x) = 1 - F_{X_{j,n}|M_{j,n}=1}(x)$. Differentiating yields the PDF
\begin{multline}
  f_{X_{j,\mathcal{A}}^{\min}|M_{j,\mathcal{A}}=1}(x) = \sum_{n\in\mathcal{A}} \Big[f_{X_{j,n}|M_{j,n}=1}(x)\\
  \cdot \prod_{\forall\,m\neq n \in\mathcal{A}} \bar{F}_{X_{j,m}|M_{j,m}=1}(x) \Big]
\end{multline}
This PDF allows the computation of~\eqref{eq:independent_mcs_rate} for an RB assignment~$\mathcal{A}$. Finally, the total rate for each MS~$j$ is the sum of the rates over all the possible assignments:
\begin{align}\label{eq:uniform_mcs_rate}
\mathcal{R}_{j} &= \sum_{\forall\,\mathcal{A} \subseteq \{1, \ldots, N \}}\mathcal{R}_{j,\mathcal{A}}\,.
\end{align}

\subsubsection{Validation} 
\begin{figure}[ht]
\centering
\includegraphics[width=3.3in]{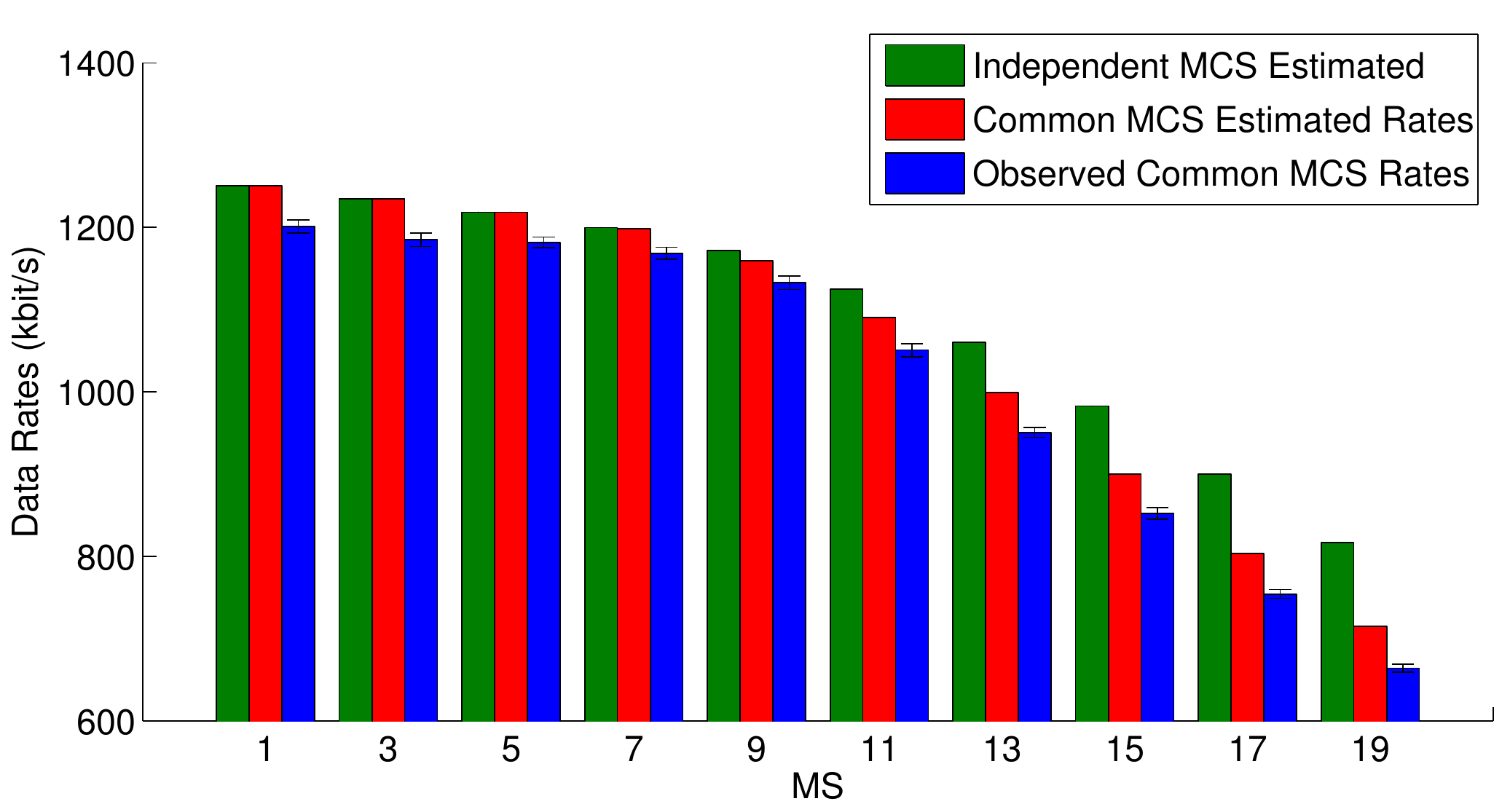}
\caption{Rate estimation comparison (model) for independent vs uniform MCS (model) vs obtained rates from simulations}
\label{fig:dyn_vs_uni_vs_sim}
\end{figure}

The same configuration parameters as in Section~\ref{section:model_evaluation_indepMCS} were used, with the exception that the jointly assigned RBs are not allowed to transmit with separate MCS anymore. Instead, the MCS selection will depend on the SINR of the worst contemporary assigned RB. The average obtained rates from simulations are plotted in Figure~\ref{fig:dyn_vs_uni_vs_sim} (observed rates) and compared against the independent and uniform MCS models. We observe a good match between the mathematical predictions and the simulated values in general where the slight discrepancy is due to the frequency correlation between resource blocks in the simulation model (which is not considered in the analytical model).

\subsection{Quantized feedback}
\begin{figure}[ht]
\centering
\includegraphics[width=3.3in]{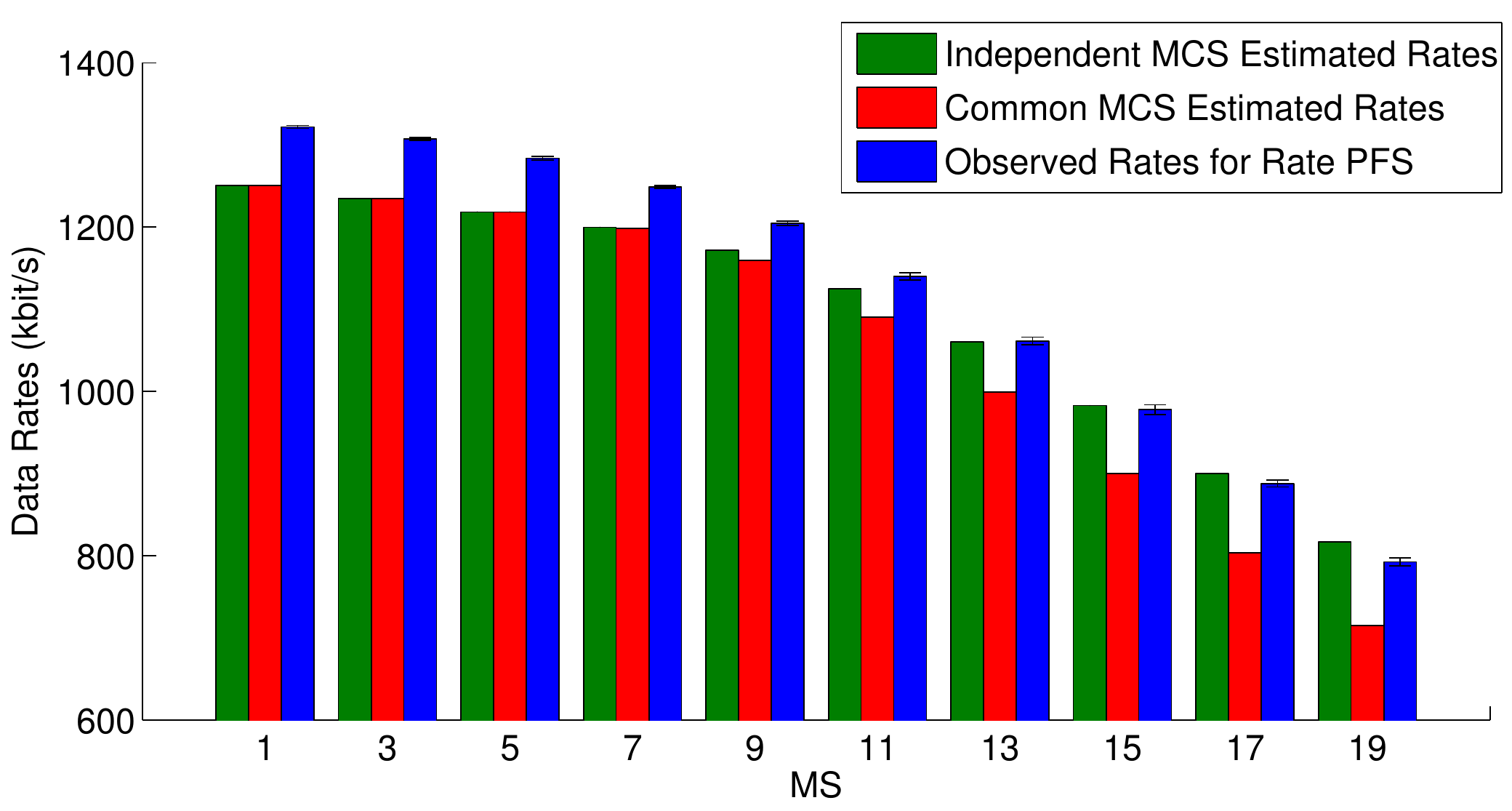}
\caption{Rate estimation comparison (models) for independent vs uniform MCS vs rate scheduler}
\label{fig:sinr_vs_rate_scheduling}
\end{figure}

A further complication comes from the fact that in LTE systems the terminals do not report the SINR to the base station. Instead, the SINR according to the rate that could be employed on the resource block is quantized. This alters the proportional fair scheduling algorithm, as they are not the normalized SINRs that are used for the scheduling decision. Instead, the algorithm considers the instantaneous rates divided by the amount of data transmitted to the terminal over that resource block during the last $W$ slots of the window. 
This 'rate-based' proportional fair scheduler is much harder to analyze precisely as the distribution of the rates has to be obtained (also taking the uniform setting of the modulation and coding schemes into account). However, we have observed that the usage of the SINR-based PFS analysis is approximately correct if also used for the rate-based PFS. To illustrate, in Figure~\ref{fig:sinr_vs_rate_scheduling} both models for independent and common MCS for SINR scheduling are compared with the obtainable rates for rate PFS. We find a slightly better performance for the simulated results from the rate-based PFS compared to the analysis for the SINR-based PFS. This vanishes for terminals closer to the cell edge.

\section{Comparison to Reference Models}
\label{sec:model_comp}

In this section we compare the results derived in this paper with several reference models from literature. The first reference model assumes instantaneous rates to be Gaussian distributed. This model is commonly encountered in related work, i.e.,~\cite{Choi2007,liu2008,liu2010,Almatarneh2010PFS,ahmed2012analytical}. In the Gaussian model, the mean $\text{E}[r_{j,n}]$ is defined as
\begin{eqnarray}
\text{E}[r_{j,n}]&=&\int_0^\infty C \left(\frac{P_{j,n}^s}{P_{j,n}^i + \eta} \cdot y \right) \cdot \exp(-y) \text{d}y 
\end{eqnarray}
and the standard deviation $\sigma_{r_{j,n}}$ is given by
\begin{eqnarray}
\sigma_{r_{j,n}}^2 &=& \int_0^\infty C\left(\frac{P_{j,n}^s}{P_{j,n}^i + \eta} \cdot y\right)^2 \cdot \exp(-y)\text{d}y \\ \nonumber 
 &&- \left( \int_0^\infty C\left(\frac{P_{j,n}^s}{P_{j,n}^i + \eta} \cdot y \right)\cdot \exp(-y) \text{d}y \right)^2 .
\end{eqnarray}
Based on the above, the average rate obtained from the PFS is calculated as follows:
\begin{eqnarray}
\mathcal{R}_{j,n} &=& \frac{R\cdot S}{T_{\text{TTI}}}\int_0^\infty \frac{\left(y \sigma_{r_{j,n}} + \text{E}[r_{j,n}]  \right)}{\sqrt{2 \pi}}  \exp\left(\frac{-y^2}{2}\right) \nonumber \\ && \cdot \prod_{i=1, i\neq j}^N \text{F}_{(0,1)} \left( \frac{\text{E}[r_{i,n}] \sigma_{r_{j,n}}}{\text{E}[r_{j,n}] \sigma_{r_{i,n}}} y \right) \text{d}y \,,
\end{eqnarray}
in which~$\text{F}_{(0,1)}(\cdot)$ is the standard Gaussian distribution function with zero mean and unit variance.

The second reference model is the "Interference as Noise" (IaN) model. Within this model, interference is regarded as an additional source of noise. Hence, the SINR is exponentially distributed according to the following PDF:
\begin{align}
f_{X_{j,n}}\left(x\right)  =  \frac{P_{j,n}^i + \eta}{P_{j,n}^s} \text{exp}\left(-\frac{P_{j,n}^i + \eta}{P_{j,n}^s} \cdot  x\right) \; \mathrm{.} \label{eqn:base_pdf_intf_as_Nosie}
\end{align}
For comparison, this function is plugged into~\eqref{eqn:transformed_SINR} and used to calculate the rates per terminal according to~\eqref{eqn:totalRateIndepMCS}.

The last reference model is what we refer to as the "Na\"ive" approach. Instead of a PDF, this model uses the simplified formula~$\bar{\gamma}_{j,n}=\frac{P_{j,n}^s}{P_{j,n}^i + \eta}$ for the SINR. For rate function~$C(\gamma)$, the resulting average rates compute to
\begin{equation}
\mathcal{R}_{j}=\sum_{n=1}^{N} \frac{R \cdot S}{J\cdot T_{\text{TTI}}} C\left( \frac{P_{j,n}^s}{P_{j,n}^i + \eta} \right).
\label{eqn:totalRateNaive}
\end{equation}

\begin{figure}[ht]
\centering
\includegraphics[width=3.3in]{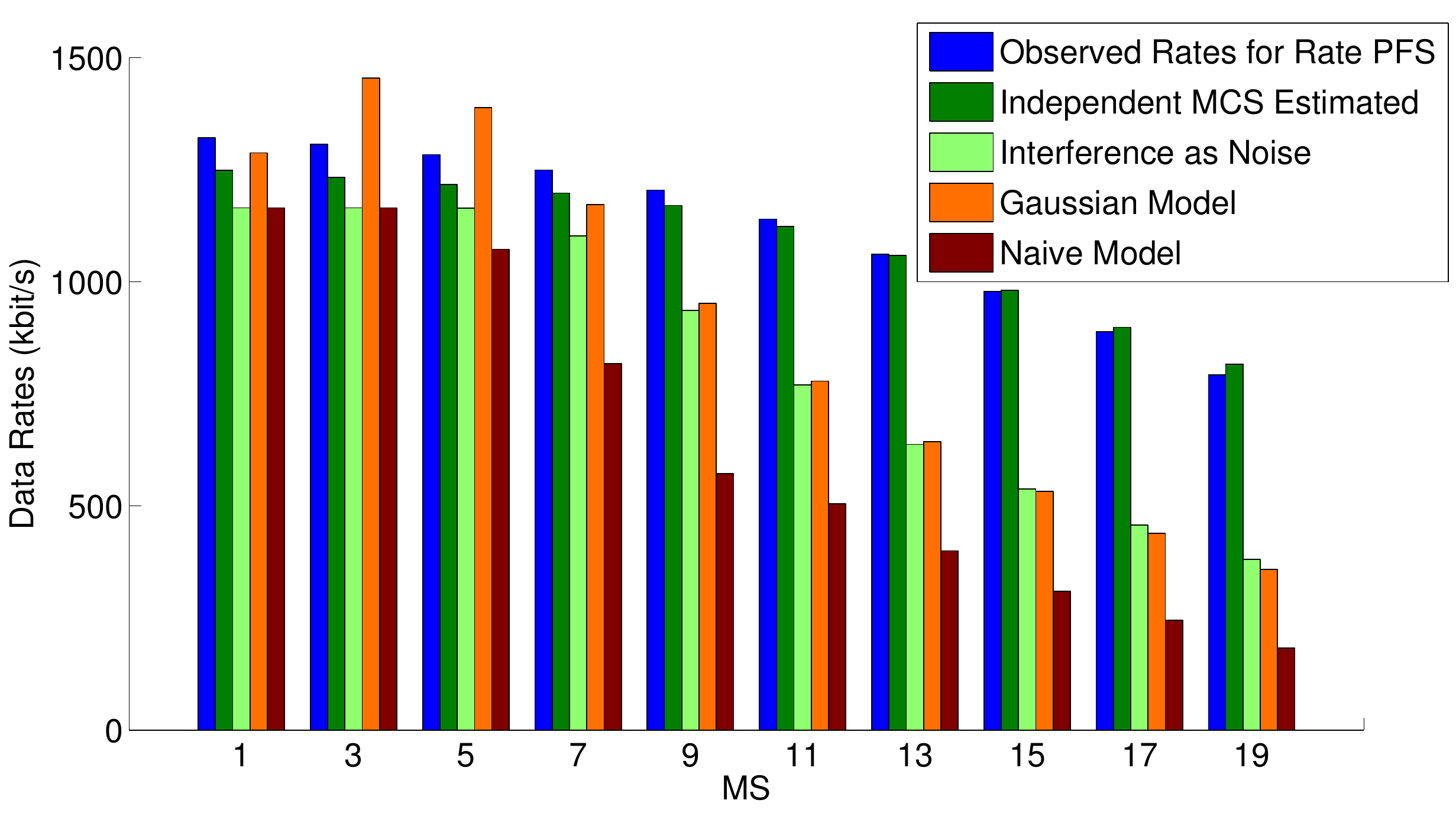}
\caption{Comparison of the observed rates from simulations with reference models and the one proposed here}
\label{fig:modell_comparison}
\end{figure}

These reference models are compared with the rates observed from simulations for the rate PFS and the ``Independent MCS'' model introduced in Formula~\ref{eq:expected_rate_rb}. The comparison is shown in Figure~\ref{fig:modell_comparison}. For mobile terminals in the cell center, the prediction quality of all schemes is comparable. However, all three reference models underestimate the data rate at the cell edge. This is particularly significant for the Na\"ive model for which the data rates decay the most towards the cell edge. However, compared with simulation observations, this is not the case. This is due to the fact that the Na\"ive model does not consider multi-user diversity gain, hence overestimating the impact of interference.

The rates predicted by the Gaussian and the IaN model are relatively similar, especially in the high interference regime of cell-edge terminals. However, just like the Na\"ive model, both severely underestimate the data rate. While Gaussian approximation works well for noise-limited systems, this shows that it is inaccurate under the presence of interference. Concluding, the three reference models overestimate the impact of interference, in particular at the cell edge. We believe that this has important implications for system tasks like interference coordination and handover decisions.

\section{Conclusions}\label{sec:conclusions}
The mathematical models introduced in this paper have been shown to be capable of describing the scheduled SINR distribution and estimating the obtainable rates in OFDMA systems employing PFS. They are valid for systems operating on independent as well as common MCS. Simulation results show that the proposed scheme with independent MCS closely approximates the data rates obtained by rate based PFS. Comparing these results to three reference schemes shows that common simplifying assumptions on the rate distribution are not applicable to interference-limited scenarios. Finally, we note that modeling of dynamic schedulers is a tedious process and needs exact knowledge of the instantaneous SINR or rate distribution.

\bibliographystyle{IEEEtran}
\bibliography{Recent_Related_Work}
%

\end{document}